\documentclass[aps,prl,twocolumn,groupedaddress]{revtex4-2}

\usepackage{amsmath}
\usepackage{amssymb}
\usepackage{amsfonts}
\usepackage[dvipsnames]{xcolor}
\usepackage{hyperref}
\hypersetup{
	pdftex,
	pdfstartview=FitH,
	bookmarksnumbered=true,
	pagebackref,
	colorlinks,
	breaklinks,
	urlcolor=Maroon,
	linkcolor=Maroon,
	citecolor=Bittersweet
} 
\usepackage{graphicx}
\DeclareGraphicsExtensions{.pdf,.png,.jpg,.mps}
\usepackage{tikz}
\usetikzlibrary{matrix,decorations.pathmorphing,decorations.markings,arrows,positioning,intersections,mindmap,backgrounds}
\usepackage{booktabs}
\usepackage{physics}
\usepackage{diagbox}
\usepackage{cleveref}

\usepackage{lmodern} 
\usepackage{scalerel,accents,trimclip}
\usepackage{CJK}

\newcommand{\zb}{\bar{z}}

\newcommand{\DU}{\mathcal{D}_{U}}
\newcommand{\DV}{\mathcal{D}_{V}}

\newcommand{\newhat}{\scalebox{2}[1.05]{\trimbox{0pt 1.15ex}{\textasciicircum}}}
\newcommand{\mhat}[1]{\accentset{\newhat}{#1}}

\usepackage{graphicx}

\begin{document}

	
	\begin{CJK*}{UTF8}{gbsn}

	\title{All Next-to-Next-to-Extremal One-Loop Correlators\\
		of AdS Supergluons and Supergravitons}


	\author{Zhongjie Huang (黄中杰)}
	\email{zjhuang@zju.edu.cn}

	\author{Bo Wang (王波)}
	\email{b\_w@zju.edu.cn}
	\author{Ellis Ye Yuan (袁野)}
	\email{eyyuan@zju.edu.cn}
	\affiliation{Zhejiang Institute of Modern Physics, School of Physics, Zhejiang University, \\Hangzhou, Zhejiang 310058, China }
	\affiliation{Joint Center for Quanta-to-Cosmos Physics, Zhejiang University,
		\\Hangzhou, Zhejiang 310058, China}

	\date{\today}

	\begin{abstract}
		We bootstrap all of the next-next-to-extremal one-loop four-point correlators of supergravitons and supergluons in ${\rm AdS_5}$ using a differential representation, and obtain closed formulas that are valid in both position space and Mellin space simultaneously.
	\end{abstract}


	\maketitle
	
	\end{CJK*}


	\noindent{\bf Introduction and Summary.}
	In studying holographic CFTs one often cares about correlation functions of the lowest half-BPS operator and its higher Kaluza-Klein (KK) modes. These observables together provide a practical approach to details of operator spectrum and interaction structure. For this purpose it is desirable to have a closed formula for all of them.
	
	This has been achieved in four-point tree-level correlators in a variety of holographic models, e.g., \cite{Rastelli:2016nze,Caron-Huot:2018kta,Giusto:2018ovt,Rastelli:2019gtj,Giusto:2019pxc,Giusto:2020neo,Alday:2020dtb,Alday:2020lbp,Alday:2021odx,Abl:2021mxo,Rigatos:2024yjo}. They can be bootstraped in either position space or Mellin space, yielding a simple and unified formula for arbitrary KK modes \footnote{The bootstrap computation at tree-level higher points has also been explored, see \cite{Goncalves:2019znr,Alday:2022lkk,Goncalves:2023oyx,Alday:2023kfm,Cao:2023cwa,Cao:2024bky}.}, which led to the observation of hidden conformal symmetries \cite{Caron-Huot:2018kta,Rastelli:2019gtj,Giusto:2019pxc,Giusto:2020neo,Abl:2021mxo,Heslop:2023gzr,Rigatos:2024yjo}. Similar closed formulas were also obtained for stringy corrections \cite{Abl:2020dbx,Aprile:2020mus,Aprile:2022tzr,Glew:2023wik}. When the analysis moves to one-loop supergravity/SYM, while this was done case by case in position space  \cite{Aprile:2017bgs,Aprile:2017qoy,Aprile:2019rep,Huang:2023oxf}, in Mellin space closed results were worked out for some restricted classes of KK configurations \cite{Alday:2018kkw,Alday:2019nin,Alday:2021ajh,Huang:2023ppy}. The analysis heavily relies on double-trace data that are available only through CFT unitarity methods \cite{Aharony:2016dwx}, which is in practice a complicated computation due to operator mixing \cite{Aprile:2017xsp,Aprile:2019rep,Alday:2021ajh,Huang:2023ppy}.
	
	In this paper, we work out closed formulas for next-next-to-extremal (N$^2$E) one-loop correlators of AdS supergluons and supergravitons. Complexity of correlators relates to the gap between the largest dimension and the sum of dimensions of the rest operators. When they are equal, the correlator is called extremal. Extremal and next-to-extremal cases are protected \cite{DHoker:1999jke,Eden:2000gg}, while N$^2$E correlator is the first non-trivial case. This class is formed by $\langle \mathcal{O}_p \mathcal{O}_q\mathcal{O}_r\mathcal{O}_{p+q+r-4}\rangle$ and their permutations, with the subscripts referring to the KK level \footnote{Our definition of extremality follows the original one in \cite{DHoker:1999jke,Eden:2000gg}. For four-point functions there exists a different definition of extremality which counts the possible R-structures (see e.g., \cite{Bissi:2022mrs}). While the latter is more frequently used in recent literature, we adopt the former one, which processes better analytic behavior in KK configuration space within a single class (see also \cite{Huang:2024dxr}). }.
	
	 We compute these N$^2$E correlators by utilizing a differential representation that was recently extended to loop levels \cite{Huang:2024dck}, where a correlator is written as differential operators acting on certain seed functions. This can be conveniently translated into Mellin space as rational functions times seed Mellin amplitudes, providing a simple connection between functions in position and Mellin space. Hence our closed-form formulas are available in both spaces \emph{simultaneously}, which greatly extends the previous results. In addition, the existence of this representation puts strong constraints on the structure of correlators. By imposing the knowledge of leading logarithmic behavior from hidden conformal symmetries only, the differential representation directly fixes the whole class of N$^2$E supergluon correlators, and leaves one single unknown coefficient in N$^2$E supergraviton correlators. This coefficient is fixed by comparing with the known result of $\langle \mathcal{O}_3 \mathcal{O}_3\mathcal{O}_3\mathcal{O}_5\rangle$ in \cite{Aprile:2019rep}.  The explicit closed-form results for supergluons are contained in \eqref{eq:FST}\eqref{eq:Hextremality}, and those for supergravitons in supplemental materials.

	\vspace{0.4em}
	\noindent{\bf Holographic models.} 
	The supergravitons live in the type IIB supergravity on $\mathrm{AdS}_5\times\mathrm{S}^5$. On the boundary they correspond to half-BPS operators $\mathcal{O}_{p\geq 2}(x;y)$, with protected dimension $\Delta=p$. These operators transform as $[0,p,0]$ of the $\mathrm{SU}(4)\simeq\mathrm{SO}(6)$ R-symmetry group, characterized by an $\mathrm{SO}(6)$ null vector $y$.

	The supergluons live on $\mathrm{AdS}_5\times\mathrm{S}^3$, which can be, e.g., realized by inserting probe D7 branes in the above model \cite{Karch:2002sh}\footnote{Another construction is to probe an F-theory D7-brane singularity using a stack of D3-branes \cite{Fayyazuddin:1998fb,Aharony:1998xz}. }. They correspond to half-BPS operators $\mathcal{O}_{p\geq 2}^I(x;v,\bar{v})$, with protected dimension $\Delta=p$. The index $I$ labels the adjoint representation of a color group $G_F$. These operators transform in the spin-$\frac{p}{2}$ representation of the $\mathrm{SU}(2)_R$ R-symmetry group, and in the spin-$\frac{p-2}{2}$ representation of an additional $\mathrm{SU}(2)_L$ flavor symmetry group, which are taken care of by the polarization spinors $v$ and $\bar{v}$ respectively. 
	
    Viewing from the boundary, both models are described by superconformal field theories, with $\mathcal{N}=4$ supersymmetries in the former and $\mathcal{N}=2$ in the latter \footnote{Rigorously speaking, the full superconformal field will be dual to a full string theory in the AdS bulk. In this letter we only focus on the low-energy limit, without considering stringy corrections. And in the supergluon case, while in principle there are also gravitons in the bulk, we only restrict to the gluon sector.}. For any four-point correlator $\langle p_1p_2p_3p_4\rangle\equiv\langle \mathcal{O}_{p_1}\mathcal{O}_{p_2}\mathcal{O}_{p_3}\mathcal{O}_{p_4}\rangle$, by solving superconformal Ward identities one can split out a universal part, so that the actual dynamics of the correlator is fully contained in a so-called reduced correlator $\mathcal{H}_{\{p_i\}}$  \cite{Eden:2000bk,Nirschl:2004pa}. For N$^2$E correlators the latter only depend on the spacetime cross ratios \footnote{Note that here we adopt a slightly different convention on the reduced correlator to factor out all the R-symmetry dependence, which is further explained in the supplemental materials.}
	\begin{align*}
		U=\frac{x_{12}^2x_{34}^2}{x_{13}^2x_{24}^2}\,,\quad
		V=\frac{x_{14}^2x_{23}^2}{x_{13}^2x_{24}^2}\,,\quad x_{ij}^2\equiv(x_i-x_j)^2.
	\end{align*}
	A convenient convention for its Mellin amplitude is
	\begin{align*}
		\mathcal{H}_{\{p_i\}}(U,V)=\!\int\!\frac{\mathrm{d}S\mathrm{d}T}{(2\pi i)^2}U^{S}V^{T}\!\mathcal{M}_{\{p_i\}}(S,T )\Gamma_{\{p_i\}}(S,T)\, ,
	\end{align*}
	where
	\begin{align}\label{eq:Gammacanonical}
		\Gamma_{\{p_i\}}(S,T)=&\ \Gamma(-S)\Gamma(-T)\Gamma(-\tilde{U}) \nonumber \\
		&\times\Gamma(\omega_s-S)\Gamma(\omega_t-T)\Gamma(\omega_u-\tilde{U})\, , 
	\end{align}
	with $S+T+\tilde{U}=-2-\frac{\mathcal{N}}{2}$, $\omega_s=\frac{|\Sigma_{12,34}| }{2}$, $\omega_t=\frac{|\Sigma_{14,23}| }{2}$, $\omega_u=\frac{|\Sigma_{13,24}| }{2}$ and $\Sigma_{ij,kl}=p_i+p_j-p_k-p_l$.  
	
	In the weak-coupling limit of the bulk theories, $\mathcal{H}$ receives a loop expansion with respect to a bulk coupling parameter $a$ \footnote{This parameter $a$ is inversely proportional to the central charge in the $\mathrm{AdS}_5\times\mathrm{S}^5$ model and the flavor central charge of $G_F$ in the $\mathrm{AdS}_5\times\mathrm{S}^3$ model, respectively.}
	\begin{align*}
	    \mathcal{H}=\mathcal{H}^{(0)}+a\,\mathcal{H}^{(1)}+a^2\mathcal{H}^{(2)}+\cdots
	\end{align*}
	and similarly for Mellin amplitudes $\mathcal{M}$. The one-loop reduced correlator we study is $\mathcal{H}^{(2)}$. In Mellin space, it can be decomposed on three channels 
	\begin{equation}\label{eq:oneloopMellin}
	    \mathcal{M}^{(2)} = \mathcal{M}^{(2)}_{st} + \mathcal{M}^{(2)}_{su} + \mathcal{M}^{(2)}_{tu}
	\end{equation}
	according to their pole structures. These channels are related by Bose symmetry. Each of them are usually written as the sum of infinitely many poles. And for all cases currently known \cite{Alday:2018kkw,Alday:2019nin,Alday:2021ajh,Huang:2023ppy}, all these poles are simultaneous poles depending on two of the Mellin variables $S,T,\tilde{U}$. Especially, for N$^2$E correlators we expect
	\begin{equation}\label{eq:ansatz}
	    \mathcal{M}^{(2)}_{st} = \mathcal{C}_{st} \sum_{m,n=0}^\infty \frac{a_{m,n}}{(S-m)(T-n)}\, ,
	\end{equation}
    where $\mathcal{C}_{st}=1$ for supergravitons, and 
    \begin{align*}
        \mathcal{C}_{st} = \parbox{1.1cm}{\tikz[every node/.style={inner sep=1pt},every path/.style={draw,very thick}]{\draw (-0.2,-0.2) rectangle (0.2,0.2);\draw (-0.2,0.2) -- (135:0.5) node [left] {\scriptsize $I_1$};\draw (-0.2,-0.2) -- (-135:0.5) node [left] {\scriptsize $I_2$};\draw (0.2,-0.2) -- (-45:0.5) node [right] {\scriptsize $I_3$};\draw (0.2,0.2) -- (45:0.5) node [right] {\scriptsize $I_4$};}} \quad \equiv f^{J\, I_1 K} f^{K\, I_2 L} f^{L\, I_3 M} f^{M\, I_4 J}
    \end{align*}
    the color structure for supergluons. The residue $a_{m,n}$ encodes the one-loop dynamics and needs to be determined.

	\vspace{0.4em}
	\noindent{\bf Differential Representation.} 
	In order to easily work with the differential representation, it is preferable to rewrite Mellin amplitudes in a standard form (indicated by hats), whose definition is independent of external operator dimensions $\{p_i\}$ 
	\begin{align}\label{eq:StandardMellinTransform}
		\mathcal{H}(U,V)\!=\!\int\! \frac{\dd S \dd T}{(2\pi i)^2} U^S V^T\, \mhat{\Gamma}(S,T)\, \mhat{\mathcal{M}}(S,T)\, ,
	\end{align}
	where 
	\begin{align*}
		\mhat{\Gamma}(S,T)\equiv\Gamma^2(-S)\Gamma^2(-T)\Gamma^2(1+S+T)\, .
	\end{align*}
	$\mhat{\mathcal{M}}$ relates to the usual Mellin amplitude $\mathcal{M}$ by 
	\begin{align}\label{eq:stdmellin}
	    \mhat{\mathcal{M}} &= \frac{\Gamma_{\{p_i\}}(S,T)}{\text{\trimbox{0pt 1.15ex}{$\mhat{\Gamma}(S,T)$}}}{\mathcal{M}}_{\{p_i\}} ,
	\end{align}
	and the $\Gamma$ ratio here is
	\begin{align}\label{eq:gammaratio}
	    (-S)_{\omega_s}(-T)_{\omega_t}(1+S+T)_{1+\frac{\mathcal{N}}{2}}(1+S+T)_{1+\frac{\mathcal{N}}{2}+\omega_u},
	\end{align}
	where $(a)_n\equiv\Gamma(a+n)/\Gamma(a)$ is the Pochhammer symbol. Since all the subscripts in \eqref{eq:gammaratio} are non-negative integers, it is in fact a polynomial. 
	
	The differential operators used in this representation are always polynomials of $U$, $U^{-1}$, $V$, $V^{-1}$, $\partial_U$ and $\partial_V$. Given the convention \eqref{eq:StandardMellinTransform}, these operators can always be decomposed into four elementary operators that acts in simple ways on both the position side and the Mellin side. These consist of two multiplication operators 
	\begin{subequations}
	\label{eq:ruleUV}
	\begin{align}
	    U^a{\mathcal{H}}(U,V) &\Leftrightarrow  \big[(-S)_{a}(1\!+\!S\!+\!T)_{-a}\big]^2 \mhat{\mathcal{M}}(S\!-\!a,T)\, , \label{eq:ruleU}\\
		V^b{\mathcal{H}}(U,V) &\Leftrightarrow  \big[(-T)_{b}(1\!+\!S\!+\!T)_{-b}\big]^2 \mhat{\mathcal{M}}(S,T\!-\!b)\, , \label{eq:ruleV}
	\end{align}
	\end{subequations}
	and two differential operators ($\mathcal{D}_x \equiv x \partial_x$)
	\begin{subequations}
	\label{eq:ruleDUV}
	\begin{align}
		(\DU)^a\,{\mathcal{H}}(U,V)\quad &\Leftrightarrow\quad S^a\mhat{\mathcal{M}}(S,T)\,, \label{eq:ruleDU}\\
		(\DV)^b\,{\mathcal{H}}(U,V)\quad &\Leftrightarrow\quad T^b\mhat{\mathcal{M}}(S,T)\,. \label{eq:ruleDV}
	\end{align}
	\end{subequations}

	Up to one loop there are three seed functions ${\mathcal{W}}_{i=2,3,4}(U,V)$ \cite{Aprile:2020luw,Huang:2024dck} \footnote{In position space, these seed functions are single-valued polylogarithms \cite{Goncharov:1998kja,Goncharov:2001iea}. See Supplemental Material for a brief review. }. Their corresponding Mellin amplitudes $\mhat{\mathcal{M}}_i$ via the relation \eqref{eq:StandardMellinTransform} are simple 
	\begin{gather*}
		\mhat{\mathcal{M}}_2(S,T)=1,\qquad  \mhat{\mathcal{M}}_3(S,T)\equiv\xi_S=\psi^{(0)}(-S)+\gamma_E\, , \\
		\mhat{\mathcal{M}}_4(S,T) \equiv \Phi_{S,T} =  -\frac{1}{2}\big[(\xi_S-\xi_T)^2+\xi'_S+\xi'_T+\pi^2\big]\, ,
	\end{gather*}
	where we abbreviate $f_X\equiv f(X)$. Explicit position space expressions of $\mathcal{W}_i$ are collected in the supplemental materials. Using these seed functions we can generally write a one-loop correlator as 
	\begin{align*}
	    \sum_{i} \mathcal{D}_i(U,V,\DU,\!\DV) {\mathcal{W}}_i\Leftrightarrow \sum_{i,a,b} F_{i,a,b}(S,T) \mhat{\mathcal{M}}_i(S\!-\!a,T\!-\!b),
	\end{align*}
	where $\mathcal{D}_i$ are differential operators and $F_{i,a,b}$ are the corresponding rational functions in Mellin space. 

	Functions $\xi_S$ and $\Phi_{S,T}$ have a simple feature that residues of their leading-order poles are uniform
	\begin{align*}
	    \underset{S=m}{\mathrm{Res}}\, \xi_S=1,\quad
	    \underset{S=m}{\mathrm{Res}}\, \underset{T=n}{\mathrm{Res}}\, \Phi_{S,T}=1,\quad
	    \forall m,n\in\mathbb{N}.
	\end{align*}
	One can interpret this structure as, e.g., $\Phi_{S,T}$ provides a grid of poles, such that the product $f(S,T)\Phi_{S,T}$ yields residues $f_{m,n}$ (as long as $f(S,T)$ is regular at these points). Given that many Mellin amplitudes in these models are known in terms of residues, one can reverse this logic and organize them into combinations of $1$, $\xi$ and $\Phi$ dressed with proper rational functions.

\vspace{0.4em}
	\noindent{\bf Analyticity of residues.} The rational functions mentioned above are also the Mellin counterpart of differential operators. Therefore, they are built from the elementary operations \eqref{eq:ruleUV}\eqref{eq:ruleDUV}, and possess the \textit{double zero property}: denominators of $F_{i,a,b}$ and shifts in $\mhat{\mathcal{M}}_i$ can only arise from the multiplication operators \eqref{eq:ruleUV}, which in turn introduces a series of double zeros in $F_{i,a,b}$. To be concrete, consider the action of $U^{a>0}V^{b>0}$. It gives 
	\begin{equation}\label{eq:doublezero}
	    \frac{\prod_{m=0}^{a-1}(S-m)^2 \prod_{n=0}^{b-1}(T-n)^2}{\prod_{l=0}^{a+b-1}(S+T-l)^2} \mhat{\mathcal{M}}_i(S-a,T-b)\, ,
	\end{equation}
	which indicates that any appearance of a factor $S+T-(a+b-1)$ in the denominator of $F_{i,a,b}$ has to be accompanied by a shift in $\mhat{\mathcal{M}}_i$ and a collection of double zeros $\prod_{m=0}^{a-1}(S-m)^2 \prod_{n=0}^{b-1}(T-n)^2$ in the numerator. Note that since the action \eqref{eq:ruleDUV} of the differentials $\mathcal{D}_U$ and $\mathcal{D}_V$ can at most create a polynomial of $S$ and $T$, they can never lower the degree of any existing zero in the numerator. 
	
	Based on this double zero property, we can show that for a correlator admitting differential representation, the corresponding residues $a_{m,n}$ in \eqref{eq:ansatz} are expressed in terms of a single analytic function of $m,n$ in the first quadrant $m,n\geq 0$. We first consider the analyticity of simultaneous pole residues in the standard amplitude $\mhat{\mathcal{M}}^{(2)}$, since it is related to the analyticity of $a_{m,n}$ in $\mathcal{M}^{(2)}$ through the relation \eqref{eq:stdmellin}. Non-analyticity may potentially arise in two different ways. One is by terms like
	\begin{align*}
	    \frac{\hat{\delta}}{(S-a)(T-b)}\quad \subset \quad \mhat{\mathcal{M}}^{(2)},
	\end{align*}
	which merely affects the residue at a single point and breaks the original smooth behavior at $m=a$, $n=b$. Another way is through terms like $f(S,T)\Phi_{S-a,T-b}$, which generates residue
	\begin{align*}
	    A_{m,n} = \begin{cases}\, f_{m,n}\,, &\quad m\geq a,\ n\geq b,\\
	    \, 0\,, &\quad\text{others}, \end{cases}
	\end{align*}
	carrying a sudden change of expression at $m=a$, $n=b$. However, when either of these non-analyticity locates in the first quadrant (i.e., $a,b\geq0$), it is in fact resolved in the context of differential representation. For the first kind of non-analyticity, since the denominator $(S-a)(T-b)$ can never be generated from the multiplication operators \eqref{eq:ruleUV}, such term is simply forbidden in the differential representation. For the second one, we can always extend the domain of $A_{m,n} = f_{m,n}$ to the whole first quadrant (as long as it is well-defined), because according to double zero property, $f_{m,n}$ is already $0$ at those additional integer points. Therefore we establish the analyticity of residues in the standard amplitude $\mhat{\mathcal{M}}^{(2)}$.
	
	The analyticity of residues $a_{m,n}$ in the original amplitude $\mathcal{M}^{(2)}$ is a bit more subtle. This is because when switching from $\mathcal{M}^{(2)}$ to $\mhat{\mathcal{M}}^{(2)}$, the $\Gamma$ ratio \eqref{eq:gammaratio} may cancel certain poles and provide extra zeros. For example, consider
	\begin{align*}
	    \frac{\omega_s \omega_t}{ST}.
	\end{align*}
	When $\omega_s=0$ or $\omega_t=0$ this term vanishes, and when $\omega_s,\omega_t>0$ the pole is canceled by $(-S)_{\omega_s}(-T)_{\omega_t}$ in factor \eqref{eq:gammaratio}, since
	\begin{subequations}\label{eq:extrazeros}
	\begin{align}
	    (-S)_{\omega_s} =&\ (-S)(-S+1)\cdots(-S+\omega_s-1)\, ,\label{eq:extrazerosS}\\
	    (-T)_{\omega_t} =&\ (-T)(-T+1)\cdots(-T+\omega_t-1)\, .
	\end{align}
	\end{subequations}
	Hence it does not violate the requirement of differential representation, even though it introduces non-analyticity to $\mathcal{M}^{(2)}$. 
	More generally, pole non-analyticity
	\begin{align}\label{eq:pole}
	    \frac{\delta}{(S-a)(T-b)}\quad\subset\quad\mathcal{M}^{(2)}
	\end{align}
	is allowed when its residue obeys
	\begin{align*}
	    \delta\ \propto\ \omega_s \cdots (\omega_s-a) \times \omega_t \cdots (\omega_t-b)\, .
	\end{align*}
	This implies that the highest possible power of $\omega$ will bound the region in which the pole non-analyticity can appear. For N$^2$E supergluons and supergravitons, as we will see, the residues are at most linearly and quadratically on $\omega$, which means that in N$^2$E supergluons there is no such non-analyticity, and in N$^2$E supergravitons there is only one possible pole non-analyticity at $(m,n)=(0,0)$.
	
	Moreover, the extra zeros in \eqref{eq:extrazeros} also affects the counting of double zeros when working with $\mathcal{M}^{(2)}$. For instance both two terms 
	\begin{align}\label{eq:phizero}
	    \frac{S^2}{S+T} \Phi_{S-1,T},\qquad \frac{\omega_s S}{S+T} \Phi_{S-1,T}
	\end{align}
	satisfy the double zero property for $\mathcal{M}^{(2)}$. When $\omega_s>0$, the na\"ive missing power of $S$ in the second term is now compensated by the extra factors \eqref{eq:extrazerosS}. And when $\omega_s=0$ this term is simply absent. Therefore, we can equivalently view $\omega_s$ as providing a single power of $S$ (and similarly for $\omega_t$ and $T$). It is also worth noting that, although the residues $\frac{\omega_s m}{m+n}$ we directly read off from the second term is ambiguous at $(m,n)=(0,0)$, the actual residue at $S,T=0$ is well-defined and is $0$. 
	
	The analyticity of residues provides a powerful tool for bootstrapping one-loop correlators, especially for the N$^2$E cases \eqref{eq:ansatz}, in which all the poles are located in the first quadrant. From the hidden conformal symmetry, we can work out the analytic expression of $a_{m,n}$ in a region corresponding to leading logarithmic singularity \cite{Huang:2023ppy}. The analyticity immediately tells us that this expression is actually correct in the whole first quadrant, up to some subtle points where $a_{m,n}$ is ambiguous like in \eqref{eq:phizero} or contains pole non-analyticity as in \eqref{eq:pole} \footnote{In principle, non-analyticity like $\frac{\omega_s}{S}\xi_{T}$ may also appear, but it can be ruled out by hidden conformal symmetry. }.

	\vspace{0.4em}
	\noindent{\bf All N$^2$E supergluons.}  We first focus on all N$^2$E supergluon correlators in $\mathrm{AdS}_5\times\mathrm{S}^3$. By hidden 8d conformal symmetry we can work out the residues in the leading log region \cite{Huang:2023ppy}
	\begin{equation*}
		a_{m,n} = N_2 \times b_{m,n} \, , \quad 
        N_2 = \frac{(\Sigma -4) }{12(\omega_s)!(\omega_t)!(\omega_u)!} \, ,
    \end{equation*}
	where $\Sigma=p_1+p_2+p_3+p_4$, and
	\begin{equation*}
		b_{m,n}= (\omega _s+1 )g_{0,1}+(\omega _t+1)g_{1,0}+(\omega_u+1) g_{1,1}
	\end{equation*}
	with $g_{i,j}$ defined as
	\begin{equation*}
		g_{i,j}=\frac{(m+i)(n+j)}{(m+n+i+j-1)_2}\, .
	\end{equation*}
	From analyticity we know this expression is valid in the whole first quadrant, up to one exception $(m,n)=(0,0)$ at which $b_{m,n}$ is ambiguous due to $g_{0,1}$ and $g_{1,0}$. Following the general expression of $b_{m,n}$, it is natural to suggest 
	\begin{equation*}
	    b_{0,0}=\alpha_1\, \omega_s  + \alpha_2 \, \omega_t + \alpha_3\, \omega_u + \alpha_4,
	\end{equation*}
	with $\alpha_i$ some unknown numerical coefficients. Since it is linear in $\omega$, there is no pole non-analyticity in the residues, and in principle $b_{0,0}$ should be fixed by differential representation. Indeed, by breaking the residue $b_{m,n}$ into pieces satisfying double zero property 
	\begin{align*}
		b_{m,n} =& -\frac{\omega_s m(m-1)+m^2}{m+n} +\frac{(\omega_u+1)(m+1)^2 }{2(m+n+2)}  \nonumber \\
		& + \frac{2\omega_s m^2 - (\omega_u-1)(m^2+m)+1}{2 (m+n+1)} + (1\leftrightarrow 3) \, ,
	\end{align*}
	we can resum the whole Mellin amplitude \eqref{eq:ansatz} following \cite{Aprile:2020luw,Alday:2021ajh,Huang:2024dck}. This gives
	\begin{align}\label{eq:FST}
		{\mathcal{M}}^{(2)}_{\mathrm{YM},st} = &\ {N}_2\times\Big[ -\frac{\omega_s S(S-1) + S^2}{S+T}\Phi_{S-1,T}  \nonumber \\
		& +\frac{2\omega_s S^2  - (\omega_u-1)(S^2+S)+1}{S+T+1}\Phi_{S,T} \nonumber \\
		& + \frac{(\omega_u+1)(S+1)^2}{2(S+T+2)}\Phi_{S+1,T} - \frac{2\omega_t+\omega_u+3}{2} \xi_{S} \nonumber \\
		& + \frac{2\alpha_1 \omega_s  + 2\alpha_2 \omega_t + (2\alpha_3-1) \omega_u + 2\alpha_4-3}{4ST} \nonumber\\
		& + (1\leftrightarrow 3) + C_0\, \Big],
	\end{align}
    where $1\leftrightarrow3$ means $(S,m,\omega_s)\leftrightarrow (T,n,\omega_t)$, and $C_0$ is an arbitrary constant due to the regularization of the one-loop divergence \cite{Alday:2021ajh}. The pole term should vanish in the differential representation, because it is linear in $\omega$ and can never be proportional to $\omega_s\omega_t$. Imposing this we find
    \begin{equation*}
	    b_{0,0}=\frac{\omega_u+3}{2}\,,
	\end{equation*}
    which is exactly the same as the result given in \cite{Huang:2023ppy}. It is worth noting that, comparing to the method in \cite{Huang:2023ppy}, we get the same result nearly for free, using hidden 8d conformal symmetry as physical input only, without any prerequisite of evaluating window OPE data!
    
	By translating the rational functions in \eqref{eq:FST} to differential operators using \eqref{eq:ruleUV} and \eqref{eq:ruleDUV}, we deduce 
	\begin{align}\label{eq:Hextremality}
		{\mathcal{H}}_{\mathrm{YM},st}^{(2)} \!=\! &\ N_2\, (-\mathcal{D}_U)_{\omega_s}(-\mathcal{D}_V)_{\omega_t}(3+\mathcal{D}_U+\mathcal{D}_V)_{\omega_u}   \nonumber\\
		& \times \Big\{\left(\mathcal{D}_{-}U + \mathcal{D}_{0} + \mathcal{D}_{+}U^{-1} \right) \mathcal{W}_4(z,\zb) \nonumber\\
		&- 1/2\,(2\omega_t+\omega_u+3)\mathcal{D}_{c}\mathcal{W}_3(z,\zb) +(1\!\leftrightarrow\! 3)  \nonumber\\
		& + C_0   \mathcal{D}_{c}\mathcal{W}_2(z,\zb)\Big\}  ,
	\end{align}
	where these operators are
	\begin{align*}
		\mathcal{D}_{-} =&\ -\Delta_{0,2} \left( \omega_s\mathcal{D}_U^{-1}(\DU-1) + 1 \right) , \\
		\mathcal{D}_{0} =&\ 1/2\, \Delta_{1,2} \left( 2\omega_s \mathcal{D}_U^2  - (\omega_u-1)(\mathcal{D}_U^2+\mathcal{D}_U)+1 \right), \\
		\mathcal{D}_{+} =&\ 1/2\,(\omega_u + 1) \Delta_{2,2} (\mathcal{D}_U+1)^4 , \\
		\mathcal{D}_{c} =&\  \left( (1+\mathcal{D}_U+\mathcal{D}_V)_2\right)^2\, ,
	\end{align*}
	with
	\begin{align*}
    \Delta_{i,j} = (i+\DU+\DV)_{j-i+1}(1+i+\DU+\DV)_{j-i} \, .
    \end{align*}
    Note that because $\mathcal{D}_U$ and $\mathcal{D}_V$ commute, we can manipulate expressions of $\{\mathcal{D}_U,\mathcal{D}_V\}$ algebraically before acting them on functions of $\{U,V\}$. Although a $\mathcal{D}_U^{-1}$ appears in $\mathcal{D}_-$, the entire differential operator remains a polynomial in $\mathcal{D}_U$. This differential representation offers a closed-form formula for all N$^2$E one-loop supergluon correlators in $\mathrm{AdS}_5\times  \mathrm{S}^3$ in position space, which is nearly impossible to achieve by position space bootstrap. 
    
    Note that we have reduced the whole N$^2$E correlator as derivatives acting purely on $\mathcal{W}_{3,4}$ (while $\mathcal{W}_2$ only appears as one-loop ambiguity and can be neglected). This is far from obvious, since in position space there are in total 16 basis functions (known as single-valued multiple polylogarithms), and now all of them are in fact the differentiations on $\mathcal{W}_{3,4}$.

	\vspace{0.4em}
	\noindent{\bf All N$^2$E supergravitons.} The N$^2$E supergraviton can also be written in a similar form as \eqref{eq:Hextremality}, with a simpler structure that $\mathcal{W}_3$ drops out after adding up all three channels in \eqref{eq:oneloopMellin}. Namely, these correlators can be expressed as differentiations purely on $\mathcal{W}_{4}$. 
	
	Following the same logic as in supergluons, we can derive the residues from the hidden 10d conformal symmetry \cite{Caron-Huot:2018kta,Alday:2019nin}
	\begin{equation*}
		a_{m,n} =  N_4 \times c_{m,n} \, , \quad 
	    N_4 = \frac{ \Sigma(\Sigma -2)}{15 (\omega_s) ! (\omega_t) ! (\omega_u)!} \, ,
	\end{equation*}
    where 
	\begin{align*}
		c_{m,n}=
		&\quad\, 2 (\omega _s+2 )  (\omega_t+2) h_{1,1}+(\omega _u+1)_2\, h_{2,2}\nonumber \\
		&+ 2 (\omega _s+2 )  (\omega_u+2) h_{1,2}+(\omega _s+1)_2\, h_{0,2} \nonumber \\
 		&+2 (\omega _t+2 )  (\omega_u+2) h_{2,1}+(\omega _t+1)_2\, h_{2,0} \, ,
 	\end{align*}
	and
	\begin{equation*}
		h_{i,j}= \frac{(m+i-1)_2 (n+j-1)_2}{(m+n+i+j-3)_3}\, .
	\end{equation*}
	The residues at $(m,n) = (0,1),(1,0),(0,0)$ are again ambiguous, and the form of $c_{m,n}$ suggests the highest power of $\omega$ to be 2. This allows the pole non-analyticity at $(m,n)=(0,0)$.
	
	Like in the supergluons cases, we can break $c_{m,n}$ into pieces and resum the Mellin amplitudes into seed functions. By requiring the vanishing of pole terms at $(0,1)$ and $(1,0)$, we find
	\begin{align*}
	    c_{0,1}=&\frac{1}{6} \left(\omega _u+5\right) \left(8 \omega _t+3 \omega _u+10\right)\, , \\
	    c_{1,0}=&\frac{1}{6} \left(\omega _u+5\right) \left(8 \omega _s+3 \omega _u+10\right)\, .
	\end{align*}
	And for the residue at $(0,0)$ we have 
	\begin{equation*}
	    c_{0,0}=\frac{2}{3} (\omega_{u}+4) (\omega_{u}+5) + \alpha_0\, {\omega_s \omega_t} \, ,
	\end{equation*}
	where $\alpha_0\, {\omega_s \omega_t}$ corresponds to the possible pole non-analyticity. It turns out that $\alpha_0=0$ by comparing with the known result $\langle 3335 \rangle$ recorded in \cite{Aprile:2019rep}. 
	
	We verify the resulting Mellin amplitude of $\langle pqr(p+q+r-4)\rangle$ for $p,q,r=2,3,...,8$ using the method in \cite{Aprile:2019rep} as an independent check (all other configurations are related by crossing symmetry). One can also show that this result correctly reduces to the $\langle 22pp \rangle$ cases considered in \cite{Alday:2019nin}. Note that these examples may involve a large window region, in which case the previous method \cite{Aprile:2019rep} becomes heavy. We provide the Mellin amplitude $\mathcal{M}^{(2)}_{{\rm GR},st}$ and the corresponding position space result in supplemental materials.

	\vspace{0.4em}
	\noindent{\bf Outlook.}
	A natural future task is to consider N$^{k\geq3}$E correlators using the method developed in this letter. In \cite{Huang:2023ppy}, N$^3$E supergluons and various examples of N$^{>3}$E are studied, where the observed analytic structure of residues can also be perfectly explained by differential representation. And we expect the analytic structure of supergraviton residues to be similar as in supergluon cases. We also conjecture that, for arbitrary KK modes, the correlators can always be written as 
	\begin{align*}
	    \mathcal{H}_{st} =&\ \mathcal{D}_{4,st} \mathcal{W}_4 + \mathcal{D}_{3,st} \mathcal{W}_3 + (\text{ambiguity}),
	\end{align*}
	where $\mathcal{D}_{i,st}$ are some differential operators, and for supergravitons terms with $\mathcal{W}_3$ cancel after adding up three channels. 
	
	The pole structure and flat space limit of the corresponding seed Mellin amplitudes $\Phi$ and $\xi$ suggests that they might be treated as ``box'' and ``bubble'' diagrams in ${\rm AdS \times S}$ background \cite{Aharony:2016dwx}. Therefore, the absence of $\xi$ may also be considered as some ``no-triangle theorem''. It would be nice to find a concrete diagrammatic interpretation for our results, and develop certain loop reduction method for AdS (e.g., see \cite{Alaverdian:2024llo}). 
	
	Recently, there are also many new developments on bootstrapping tree-level higher-point correlators \cite{Goncalves:2019znr,Alday:2022lkk,Goncalves:2023oyx,Alday:2023kfm,Cao:2023cwa,Cao:2024bky} and defect correlators \cite{Chen:2023yvw,Chen:2024orp}. It is also interesting to study whether similar analytic structure continues to hold in these cases. 
	
	\vspace{0.4em}
	\begin{acknowledgments}
		\noindent{\bf Acknowledgments.}
		The authors would like to thank Paul Heslop, Arthur Lipstein, Hynek Paul, Michele Santagata, Xinan Zhou for useful discussions, and Paul Heslop, Hynek Paul, Xinan Zhou for valuable suggestions on the first draft. ZH, BW and EYY are supported by National Science Foundation of China under Grant No.~12175197 and Grand No.~12347103. EYY is also supported by National Science Foundation of China under Grant No.~11935013, and by the Fundamental Research Funds for the Chinese Central Universities under Grant No.~226-2022-00216. The authors thank Riken iTHEMS and the Yukawa Institute for Theoretical Physics at Kyoto University, where this work was at its final stage during ``iTHEMS-YITP Workshop: Bootstrap, Localization and Holography''.
	\end{acknowledgments}

	\bibliography{refs}

    	\widetext
	\begin{center}
		\textbf{\large Supplemental Materials}
	\end{center}
	\begin{appendix}

		\section{Superconformal symmetries and reduced correlators}
		
		For readers convenience, we briefly describe the reduced correlators in relation to the full four-point correlators.
		
		In the $\mathrm{AdS}_5\times \mathrm{S}^3$ case, the half-BPS operator $\mathcal{O}_p^{I}(x;v,\bar{v})$ for the supergluons and their higher KK modes is labeled by its dimension $p$, an adjoint index $I$ for its gauge group $G_F$, and two spinors $v$ and $\bar{v}$ for its polarizations in the $\mathrm{SU}(2)_{\rm R}$ R-symmetry and $\mathrm{SU}(2)_{\rm L}$ flavor symmetry respectively. Apart from the usual conformal cross ratios
		\begin{align*}
			U=\frac{x_{12}^2x_{34}^2}{x_{13}^2x_{24}^2}\equiv z\bar{z}\,,\qquad
			V=\frac{x_{14}^2x_{23}^2}{x_{13}^2x_{24}^2}\equiv(1-z)(1-\bar{z})\, ,
		\end{align*}
		one can also conveniently define cross ratios for the $\mathrm{SU}(2)_{\rm R}$ and $\mathrm{SU}(2)_{\rm L}$ internal symmetries
		\begin{align*}
			\alpha=\frac{v_{12}v_{34}}{v_{13}v_{24}},\qquad
			\beta=\frac{\bar{v}_{12}\bar{v}_{34}}{\bar{v}_{13}\bar{v}_{24}},
		\end{align*}
		where $v_{ij}\equiv\epsilon_{ab}v_i^av_j^b$, $\bar{v}_{ij}\equiv\epsilon_{ab}\bar{v}_i^a\bar{v}_j^b$ are invariant products of $\mathrm{SU}(2)$ spinors. By imposing the superconformal Ward identities \cite{Nirschl:2004pa} the correlators can be decomposed as
		\begin{align*}
			\langle &p_1p_2p_3p_4\rangle_{\rm YM}=g_{12}^\frac{\Sigma_{12}}{2}g_{34}^{\frac{\Sigma_{34}}{2}}\left(\frac{g_{24}}{g_{14}}\right)^{\frac{p_{21}}{2}}\left(\frac{g_{13}}{g_{14}}\right)^{\frac{p_{34}}{2}}\frac{1}{\bar{v}_{12}^2\bar{v}_{34}^2} \left(\mathcal{G}_{\text{YM},\{p_i\}}(z,\bar{z},\alpha,\beta)+\mathcal{R}^{(2)}\widetilde{\mathcal{H}}_{\text{YM},\{p_i\}}(z,\bar{z},\alpha,\beta)\right)\, ,
		\end{align*}
		where we omit color indices for clarity of the expression, and abbreviate $\Sigma_{ij}\equiv p_i+p_j$, $p_{ij}\equiv p_i-p_j$, $g_{ij}\equiv v_{ij}\bar{v}_{ij}/x_{ij}^2$. The factor $\mathcal{R}^{(2)}$ is determined as 
		\begin{align*}
			\mathcal{R}^{(2)}=\frac{(z-\alpha)(\bar{z}-\alpha)}{z\bar{z}\alpha^2}\,.
		\end{align*}
		Note that $\mathcal{G}_{\text{YM},\{p_i\}}$ is protected by superconformal symmetries and can be derived in the free theory limit. Hence all information about dynamics of the theory is contained in the reduced correlator $\widetilde{\mathcal{H}}_{\text{YM},\{p_i\}}$, which provides an equivalent description of the full correlator $\langle p_1p_2p_3p_4\rangle$.

		In the $\mathrm{AdS}_5\times \mathrm{S}^5$ case, the half-BPS operator $\mathcal{O}_p(x,y)$ for the supergravitons and their higher KK modes is labeled by its dimension $p$, and an $\mathrm{SO}(6)$ null vector for its polarization in the $\mathrm{SU}(4)\simeq\mathrm{SO}(6)$ R-symmetry. One can similarly introduce cross ratios for the R-symmetry
		\begin{align*}
			\sigma=\frac{y_{12}^2y_{34}^2}{y_{13}^2y_{24}^2}\equiv \alpha\bar{\alpha}\,,\qquad
			\tau=\frac{y_{14}^2y_{23}^2}{y_{13}^2y_{24}^2}\equiv (1-\alpha)(1-\bar{\alpha})\,.
		\end{align*}
		Again solving the supersymmetric Ward identities \cite{Eden:2000bk,Nirschl:2004pa} on the four-point correlator yields (this time we assign $g_{ij}\equiv y_{ij}^2/x_{ij}^2$)
		\begin{align*}
			\langle & p_1p_2p_3p_4\rangle_{\rm GR}=g_{12}^\frac{\Sigma_{12}}{2}g_{34}^{\frac{\Sigma_{34}}{2}}\left(\frac{g_{24}}{g_{14}}\right)^{\frac{p_{21}}{2}}\left(\frac{g_{13}}{g_{14}}\right)^{\frac{p_{34}}{2}}  \left(\mathcal{G}_{\text{GR},\{p_i\}}(z,\bar{z},\alpha,\bar{\alpha})+\mathcal{R}^{(4)}\widetilde{\mathcal{H}}_{\text{GR},\{p_i\}}(z,\bar{z},\alpha,\bar{\alpha})\right),
		\end{align*}
		where the factor $\mathcal{R}^{(4)}$ is determined as
		\begin{align*}
			\mathcal{R}^{(4)}=\frac{(z-\alpha)(z-\bar{\alpha})(\bar{z}-\alpha)(\bar{z}-\bar{\alpha})}{(z\bar{z}\alpha\bar{\alpha})^2}\, .
		\end{align*}
		In analogy to the previous case, $\mathcal{G}_{\text{GR},\{p_i\}}$ is protected by the supersymmetric non-renormalization theorem, and so the dynamics is fully controlled by the reduced correlator $\widetilde{\mathcal{H}}_{\text{GR},\{p_i\}}$.
		
        For N$^2$E correlators, the reduced correlators further simplifies such that all their dependence on the internal symmetry structures cleanly factorize out of the function. For supergluons we have
        \begin{align*}
            \widetilde{\mathcal{H}}_{\text{YM},\{p_i\}}(z,\bar{z},\alpha,\beta)=(\alpha \beta)^{-\frac{\omega_t+\omega_u}{2}} [(1-\alpha )(1-\beta)]^{-T_0}U^{S_0}V^{T_0}\mathcal{H}_{\text{YM},\{p_i\}}(z,\bar{z})\,,
        \end{align*}
        with $S_0\equiv\min(\frac{\Sigma_{12}}{2},\frac{\Sigma_{34}}{2})+\frac{\mathcal{N}}{2}$, $T_0\equiv\min(\frac{\Sigma_{14,23}}{2},0)$, whereas for supergravitons the same relation holds, but with $\beta$ substituted by $\bar{\alpha}$. Due to this, we simply ignore the prefactor above, and (with a slight violation of usual terminology) we refer to $\mathcal{H}_{\{p_i\}}(z,\bar{z})$ as the reduced correlator studied in this letter. 
        
		\section{Seed functions in position space}

		Now we present the position space expression of seed functions $\mathcal{W}_2$, $\mathcal{W}_3$ and $\mathcal{W}_4$, which are dual to their Mellin space counterpart $\mhat{\mathcal{M}}_2$, $\mhat{\mathcal{M}}_3$ and $\mhat{\mathcal{M}}_4$, via the inverse Mellin integral. By rewriting
		\begin{align*}
		    W_{i}(z,\zb)=(z-\zb) \mathcal{W}_{i}(z,\zb)\,,
		\end{align*}
		these seed functions become linear combinations of products of the so-called multiple polylogarithms (MPLs). The MPLs $G_{\vec{a}}(z)$ are labeled by a weight vector $\vec{a}\equiv(a_1,a_2,\ldots,a_n)$, whose size $n$ is referred to as their transcendental weight. They can be recursively defined in terms of iterated integrals \cite{Goncharov:1998kja,Goncharov:2001iea}
		\begin{align*}
			G_{a_1a_2\ldots a_n,z}\equiv G_{a_1a_2\ldots a_n}(z)=\int_0^z\frac{\mathrm{d}t}{t-a_1}G_{a_2a_3\ldots a_n}(t)\,, 
		\end{align*}
		with the special cases (where $\vec{0}_n\equiv(0,0,\ldots,0)$ is an $n$-dimensional zero vector)
		\begin{align*}
		    G(z)=1\,,\qquad
		    G_{\vec{0}_n}(z)=\frac{1}{n!}\log^nz\,.
		\end{align*}
		As some simple examples, we have $G_{0,z}=\log(z)$, $G_{1,z}=\log(1-z)$, $G_{01,z}=\mathrm{Li}_2(z)$. Hence the MPLs are generalizations of classical polylogarithms. With these elementary functions, $W_2$, $W_3$ and $W_4$ can be expressed as
		\begin{align}
			W_{2}(z,\zb)=&\ G_{0,\zb}G_{1,z}+G_{01,\zb}+G_{10,z}-(z\leftrightarrow \zb)\,, \\
			W_{3}(z,\zb)=&\  G_{00,\zb} G_{1,z}-G_{1,\zb} G_{00,z}+G_{01,\zb} G_{0,z}-2 G_{01,\zb} G_{\zb ,z}+G_{0,\zb} G_{01,z} -G_{10,\zb} G_{0,z}-G_{10,\zb} G_{1,z}+2 G_{10,\zb} G_{\zb ,z}\nonumber \\
			&+G_{0,\zb} G_{10,z}-G_{1,\zb} G_{10,z} +2 G_{1,\zb} G_{\zb 0,z}-2 G_{0,\zb} G_{\zb 1,z}+2 G_{\zb 01,z}-2 G_{\zb 10,z}-G_{001,\zb}+G_{010,\zb} -G_{100,\zb}\nonumber\\
			&+G_{101,\zb}-G_{001,z}+G_{010,z}+G_{100,z}-G_{101,z}\, , \\
			W_{4}(z,\zb)=&\ G_{01,\zb} G_{00,z}+G_{10,\zb} G_{01,z}+G_{11,\zb} G_{10,z}+G_{00,\zb} G_{11,z} +G_{001,\zb} G_{1,z}+G_{0,\zb} G_{001,z}+G_{010,\zb} G_{0,z}+G_{1,\zb} G_{010,z} \nonumber \\
			& +G_{101,\zb} G_{0,z}+G_{1,\zb} G_{101,z}+G_{110,\zb} G_{1, z } +G_{0,\zb } G_{110,z}+G_{0011,\zb}+G_{0100,\zb}+G_{1010,\zb}+G_{1101,\zb} \nonumber \\
			&+G_{0010,z}+G_{0101,z}+G_{1011,z}+G_{1100,z} - (z\leftrightarrow \zb) \, .
		\end{align}
		Although not transparent in the above form, these functions are single-valued on the Euclidean slice $\bar{z}=z^*$, as can be expected from the inverse Mellin transformation. This guarantees that any functions generated by acting differential on them are also single-valued. In addition, each $W_i$ has a uniform transcendental weight, which is indicated by its subscript $i$. 

		\section{Results of supergravitons}
		We now present the expression for all N$^2$E Mellin amplitudes of supergravitons in $\mathrm{AdS}_5\times\mathrm{S}^5$. Using the Mellin amplitudes of seed functions, their explicit results can be decomposed as
		\begin{align}\label{eq:gravitonMellinclosed}
			{\mathcal{M}}^{(2)}_{\mathrm{GR},st}=N_4 \times \left[ \sum_{i,j=-2}^2 P_{i,j}(S,T) \Phi_{S+i,T+j} + Q(S,T)\xi_S +(1 \leftrightarrow 3)\, \right],
		\end{align}
		where $P_{i,j}(S,T)$ are rational functions on $S$ and $T$
		\begin{align*}
			P_{2,0}=& \frac{\left(\omega _u+1\right) \left(\omega _u+2\right)(S+1)^2 (S+2)^2 }{4 (S+T+3)}\,,
		\end{align*}
		\vspace{-1em}
		\begin{align*}
			P_{-2,0}= (\omega_{s}-1) \omega_{s}\frac{ (S-1) S }{S+T-1} \,,
		\end{align*}
		\vspace{-1em}
		\begin{align*}
			P_{1,0}=&\omega _s \left(-\frac{(S+1)^2 \left(20 S^2-61 S-14 T^2+28 T-32\right)}{4 (S+T+2)}+\frac{\omega _t(S+1)^2 \left(S T+13 S+T^2-3 T+8\right) }{2 (S+T+2)}\right. \nonumber \\
			&\left. \ +\frac{\omega _u\left(S^2-11 S-17\right) (S+1)^2 }{S+T+2} \right) +\omega _t \left(\frac{\omega _u(S+1)^2 (6 S+17 T+29) }{S+T+2}-\frac{(S+1)^2 \left(60 S-14 T^2+205 T+264\right)}{4 (S+T+2)}\right) \nonumber \\
			&-\frac{ \omega _u(S+1)^2 \left(3 S^2-26 S-4 T^2+8 T-24\right)}{2 (S+T+2)} -\frac{(S+1)^2 \left(6 S^2-23 S-12 T^2+24 T-3\right)}{2 (S+T+2)} \nonumber\\
			& -\frac{\omega _s^2 (S+1)^2  \left(2 S^2-9 S-2 T^2\right)}{4 (S+T+2)}-\frac{\omega _u^2 (S+1)^2 \left(3 S^2+3 S+T-2\right) }{6 (S+T+2)}  -\frac{\omega _t^2(S+1)^2 (17 T+8) }{4 (S+T+2)}  \, , 
		\end{align*}
		\vspace{-1em}
		\begin{align*}
			P_{0,0}=& -\frac{\omega _s \omega _t \left(S^3 T+13 S^3+20 S^2+15 S+4\right) }{2 (S+T+1)} +\frac{-36 S^2 T^2+18 S^4-34 S^3-38 S^2+61 S T+40 S+21}{4 (S+T+1)} \nonumber\\
			&-\frac{\omega _s  \omega _u\left(2 S^4-11 S^3-30 S^2+17 S T^2+23 S T-11 S+6 T^3+35 T^2+35 T+6\right)}{S+T+1} \nonumber\\
			&+\frac{\omega _s^2 \left(-2 S^2 T^2+4 S^4-13 S^3-7 S^2+13 S T^2+21 S T+6 S+6 T^2+10 T+4\right)}{4 (S+T+1)} \nonumber\\
			&+\frac{\omega _s \left(-42 S^2 T^2+40 S^4-29 S^3-63 S^2+177 S T^2+349 S T+98 S+60 T^3+310 T^2+366 T+116\right)}{4 (S+T+1)} \nonumber\\
			&+\frac{\omega _u^2\left(2 S^2 T+3 S^4-13 S^2+2 S T^2-14 S+3 T^4-13 T^2-14 T-4\right) }{24 (S+T+1)} \nonumber\\
			&+\frac{\omega _u\left(-32 S^2 T^2+3 S^4-58 S^3-115 S^2-86 S+3 T^4-58 T^3-115 T^2-86 T-32\right) }{8 (S+T+1)}  \, , 
		\end{align*}
		\vspace{-1em}
		\begin{align*}
			P_{-1,0}=&\omega _s \left(-\frac{S \left(5 S^3-2 S^2+7 S-6 T^2-6 T-10\right)}{S+T}-\frac{S \left(S T^2+S+2 T\right) \omega _t}{S+T}+\frac{(S-2) (S-1) S (S+1) \omega _u}{S+T}\right) \nonumber \\
			&-\frac{S \omega _s^2 \left(S^3-4 S^2+6 S-T^3-4 T^2-4 T-4\right)}{2 (S+T)}+\frac{2 S^2 (T-1) (T+1) \omega _u}{S+T}-\frac{2 \left(S^2+5\right) S^2}{S+T}  \, , 
		\end{align*}
		\vspace{-1em}
		\begin{align*}
			P_{-1,-1}=&- \omega _s^2\frac{S T^2(T+3) }{2 (S+T-1)}+ \omega _s\frac{S T^2 (7 S-12) }{2 (S+T-1)} +\frac{3 S^2 T^2}{S+T-1}+\omega _s \omega _t \frac{S T (S T+2)}{2 (S+T-1)}  \, ,
		\end{align*}
		while all other $P_{i,j}$ are $0$. On the other hand $Q(S,T)$ is a polynomial on $S$ and $T$
		\begin{align*}
			Q(S,T)=&+ \frac{1}{4} S \left(\omega _s \left(20 \omega _t-44 \omega _u+37\right)+\omega _s^2+4 \omega _t \left(5 \omega _u-24\right)-4 \omega _t^2+\omega _u^2+37 \omega _u-20\right) \nonumber\\
			&+\frac{1}{4} T \left(\omega _s \left(-24 \omega _t-24 \omega _u+74\right)+2 \omega _s^2+\omega _t \left(40 \omega _u-59\right)-3 \omega _t^2-3 \omega _u^2-59 \omega _u-40\right) \nonumber\\
			&-\omega _s \left(\omega _t+23 \omega _u-37\right)+\omega _s^2+10 \omega _t \left(2 \omega _u-5\right)-2 \omega _t^2-\omega _u^2-9 \omega _u-20  \, .
		\end{align*}
		All of these rational functions satisfy the double zero property hence this allows us to transform them as differential operators. It is straightforward to write down the differential representation in position space 
		\begin{align}\label{eq:gravitonpositionclosed}
			{\mathcal{H}}^{(2)}_{\mathrm{GR},st}=N_4 \times \left[ \sum_{i,j=-2}^{2} \mathcal{P}_{i,j}(\DU,\DV,U,V) \mathcal{W}_{4}(z,\zb) + \mathcal{Q}(\DU,\DV)  \mathcal{W}_{3}(z,\zb)+(1 \leftrightarrow 3)\, \right],
		\end{align}
		where $\mathcal{P}_{i,j}(\DU,\DV,U,V)$ and $\mathcal{Q}(\DU,\DV)$ are differential operators with polynomial dependence on $U$, $U^{-1}$, $V$, $V^{-1}$, $\DU$ and $\DU$. Their explicit expressions are
		\begin{align*}
		    \mathcal{P}_{2,0}=  \frac{1}{4} \left(\omega _u+1\right) \left(\omega _u+2\right)(\DU+1)^4 (\DU+2)^4 \Delta_{3,3} (-\DU)_{\omega _s} (-\DV)_{\omega _t} (\DU+\DV+4)_{\omega _u} U^{-2} \, ,
		\end{align*}
		\vspace{-1.2em}
		\begin{align*}
		    \mathcal{P}_{1,0}=& (\DU+1)^4 \left[\frac{1}{4} \left(-2 \DU^2+9 \DU+2 \DV^2\right) \omega _s^2+\frac{1}{4} \left(-20 \DU^2+61 \DU+14 \DV^2-28 \DV+32\right) \omega _s\right. \nonumber \\
		    &+\frac{1}{2} \left(-3 \DU^2+26 \DU+4 \DV^2-8 \DV+24\right) \omega _u+\frac{1}{2} \left(-6 \DU^2+23 \DU+12 \DV^2-24 \DV+3\right)  \nonumber  \\
		    &+\frac{1}{6} \left(-3 \DU^2-3 \DU-\DV+2\right) \omega _u^2+\left(\DU^2-11 \DU-17\right) \omega _s \omega _u  \nonumber \\
		    &+\frac{1}{2} \left(\DU \DV+13 \DU+\DV^2-3 \DV+8\right) \omega _s \omega _t+\frac{1}{4} \left(-60 \DU+14 \DV^2-205 \DV-264\right) \omega _t  \nonumber \\
		    &\left. +(6 \DU+17 \DV+29) \omega _t \omega _u+\frac{1}{4} (-17 \DV-8) \omega _t^2 \right]\Delta_{2,3}(-\DU)_{\omega _s} (-\DV)_{\omega _t} (\DU+\DV+4)_{\omega _u} U^{-1} \, ,
		\end{align*}
		\vspace{-1.2em}
		\begin{align*}
		    \mathcal{P}_{0,0}= & \bigg[ \left(-2 \DU^4+11 \DU^3+30 \DU^2-17 \DU \DV^2-23 \DU \DV+11 \DU-6 \DV^3-35 \DV^2-35 \DV-6\right) \omega _s \omega _u  \nonumber  \\
		    &+\frac{1}{4} \left(-2 \DU^2 \DV^2+4 \DU^4-13 \DU^3-7 \DU^2+13 \DU \DV^2+21 \DU \DV+6 \DU+6 \DV^2+10 \DV+4\right) \omega _s^2\nonumber  \\
		    &+\frac{1}{4} \left(-42 \DU^2 \DV^2+40 \DU^4-29 \DU^3-63 \DU^2+177 \DU \DV^2+349 \DU \DV+98 \DU+60 \DV^3+310 \DV^2+366 \DV+116\right) \omega _s \nonumber  \\
		    &+\frac{1}{24} \left(2 \DU^2 \DV+3 \DU^4-13 \DU^2+2 \DU \DV^2-14 \DU+3 \DV^4-13 \DV^2-14 \DV-4\right) \omega _u^2\nonumber  \\
		    &+\frac{1}{8} \left(-32 \DU^2 \DV^2+3 \DU^4-58 \DU^3-115 \DU^2-86 \DU+3 \DV^4-58 \DV^3-115 \DV^2-86 \DV-32\right) \omega _u\nonumber  \\
		    &+\frac{1}{4} \left(-36 \DU^2 \DV^2+18 \DU^4-34 \DU^3-38 \DU^2+61 \DU \DV+40 \DU+21\right)\nonumber  \\
		    &\left. +\frac{1}{2} \left(-\DU^3 \DV-13 \DU^3-20 \DU^2-15 \DU-4\right) \omega _s \omega _t \right]\Delta_{1,3}(-\DU)_{\omega _s} (-\DV)_{\omega _t} (\DU+\DV+4)_{\omega _u}\, ,
		\end{align*}
		\vspace{-1.2em}
		\begin{align*}
		    \mathcal{P}_{-1,0}= &\left[  \frac{1}{2} \left(\DU^3-4 \DU^2+6 \DU-\DV^3-4 \DV^2-4 \DV-4\right) \omega _s^2+5 \DU^3 \omega _s-2 \DU^2 \omega _s+2 \left(\DU^2+5\right) \DU \right. \nonumber \\
		    &+\left(\DU \DV^2+\DU+2 \DV\right) \omega _s \omega _t-2 \DU (\DV-1) (\DV+1) \omega _u-(\DU-2) (\DU-1) (\DU+1) \omega _s \omega_u \nonumber \\
		    & \left. +7 \DU \omega _s-6 \DV^2 \omega _s-6 \DV \omega _s-10 \omega _s \vphantom{\frac{1}{1}}\right ]\Delta_{0,3}(1-\DU)_{\omega _s-1} (-\DV)_{\omega _t} (\DU+\DV+4)_{\omega _u} U\, ,
		\end{align*}
		\vspace{-1.2em}
		\begin{align*}
		    \mathcal{P}_{-2,0}=(\omega_{s}-1) \omega_{s} \Delta_{-1,3} (-\DU+2)_{\omega _s-2} (-\DV)_{\omega_t}  (\DU+\DV+4)_{\omega_u} U^2 \, ,
		\end{align*}
		\vspace{-1.2em}
		\begin{align*}
		     \mathcal{P}_{-1,-1}= &  (-\DU+1)_{\omega_s-1} (-\DV+1)_{\omega_t-1} (\DU+\DV+4)_{\omega _u} \Delta_{-1,3} \nonumber  \\
		\quad &\times \left[ -\frac{\omega_{s}^2}{2} (\DV+3)  \DV  +\frac{\omega_{s}}{2} (7 \DU-12) \DV    +3 \DU  \DV  \ +\omega_s \omega_t (\DU \DV+2) \right]  U V \, ,
		\end{align*}
        \vspace{-1.2em}
		\begin{align*}
		    \mathcal{Q}(\DU,\DV) = &\; (1+\DU+\DV)^2 (2+\DU+\DV)^2 (3+\DU+\DV)^2 (-\DU)_{\omega _s} (-\DV)_{\omega _t} (\DU+\DV+4)_{\omega _u}  \nonumber \\
		    &\times \left[\hspace{5pt} \frac{1}{4} \DU \left(20 \omega _s \omega _t-44 \omega _s \omega _u+\omega _s^2+37 \omega _s+20 \omega _t \omega _u-4 \omega _t^2-96 \omega _t+\omega _u^2+37 \omega _u-20\right) \right. \nonumber \\
		    &\quad +\frac{1}{4} \DV \left(-24 \omega _s \omega _t-24 \omega _s \omega _u+2 \omega _s^2+74 \omega _s+40 \omega _t \omega _u-3 \omega _t^2-59 \omega _t-3 \omega _u^2-59 \omega _u-40\right)\nonumber \\
		    &\quad \left. -\omega _s \omega _t-23 \omega _s \omega _u+\omega _s^2+37 \omega _s+20 \omega _t \omega _u-2 \omega _t^2-50 \omega _t-\omega _u^2-9 \omega _u-20 \vphantom{\frac{1}{1}}  \right ].
		\end{align*}
    For readers' convenience, all these formulas are also collected in Mathematica notebook in the ancillary file on arXiv.

	\end{appendix}
    
\end{document}